\newif\ifjournal\journalfalse

\ifjournal
\documentclass[pop,aip,preprint,amsmath,amssymb]{revtex4-1}
\else
\documentclass[pop,aip,preprint,amsmath,amssymb]{revtex4-1}
\fi

\usepackage{graphicx}
\usepackage{bm}
\usepackage{color}
\renewcommand{\vec}[1]{\boldsymbol{#1}}

\renewcommand{\div}{\nabla \cdot}
\newcommand{\rot}{\nabla\times}

\def\aap{{\itshape Astron. Astrophys.}}
\def\apj{{\itshape Astrophys. J.}}
\def\grl{{\itshape Geophys. Res. Lett.}}
\def\jcp{{\itshape J. Comput. Phys.}}
\def\jgr{{\itshape J. Geophys. Res.}}

\def\pasj{{\itshape Publ. Astron. Soc. Japan}}
\def\pop{{\itshape Phys. Plasmas}}
\def\prl{{\itshape Phys. Rev. Lett.}}
\def\ssr{{\itshape Space Sci. Rev.}}

\begin{document}

\title[PHYSICS OF PLASMAS]{Magnetohydrodynamic structure of a plasmoid
in fast reconnection in low-beta plasmas}

\author{Seiji Zenitani}
\email{Seiji.Zenitani-1@nasa.gov}
\affiliation{
NASA Goddard Space Flight Center, Greenbelt, MD 20771, USA
}

\author{Takahiro Miyoshi}
\affiliation{
Department of Physical Science, Graduate School of Science, Hiroshima University, Higashi-Hiroshima 739-8526, Japan
}

\date{Submitted 19 November 2010; accepted 10 January 2011; published online 16 February 2011}

\begin{abstract}
Plasmoid structures in fast reconnection in low-beta plasmas
are investigated by two-dimensional magnetohydrodynamic simulations.
A high-resolution shock-capturing code enables us
to explore a variety of shock structures:
vertical slow shocks behind the plasmoid,
another slow shocks in the outer-region, and
the shock-reflection in the front side.
The Kelvin-Helmholtz-like turbulence is also found inside the plasmoid. 
It is concluded that these shocks are rigorous features
in reconnection in low-beta plasmas,
where the reconnection jet speed or the upstream Alfv\'{e}n speed
exceeds the sound speed.
\end{abstract}

\maketitle

\section{Introduction}

Magnetic reconnection \cite{sweet,parker,petschek} is a fundamental mechanism to power
various events in laboratory, space, solar, and astrophysical plasmas.
The reconnection process releases a fast plasma jet
by consuming the magnetic energy stored in the upstream region.
Such a plasma jet further interacts with the external environments and
exhibits complex plasma phenomena. 
One of the most characteristic phenomena is a ``plasmoid'',
a magnetic island with plasmas embedded inside the magnetic loop. 
Plasmoids are well accepted to explain 
the satellite observation in the Earth's magnetotail
\cite{hones77,slavin84,slavin03} and
morphological features in solar flare
and coronal mass ejections \cite{shibata95,lin08,linton09}.

Motivated by the above and many other important reasons,
the nonlinear evolution of magnetic reconnection and
the associated plasmoid evolution has been studied over many decades.
Significant insights have been obtained by
magnetohydrodynamic (MHD) simulations in
two \cite{ugai77,forbes83,ugai86,ugai87,scho89,ugai92,ugai95,nitta01,shuei01,bis01,tanuma07,barta08,murphy10,yu11}
and three dimensions \cite{birn81,hesse91,ugai91,ugai96,ugai08}.
These simulations showed that
the nonlinear evolution of the reconnection-plasmoid system
involves complicated structures, such as
slow shocks of the Petschek outflow \cite{petschek} and 
various shocks around the plasmoid \cite{forbes83,ugai87,ugai95,shuei01}. 

Recently, reconnection research has been rapidly developing
in relativistic astrophysics, too.\cite{zeni05b,naoyuki06,zeni09a,zeni10b}
In this context \citet{zeni10b} carried out
MHD simulations on relativistic magnetic reconnection.
They discussed several shock structures
that were not known in the nonrelativistic plasmoid physics:
the postplasmoid vertical shocks and
the shock-reflection structure in the plasmoid-front.

It is known that reconnection is
highly influenced by upstream conditions.
One key parameter is the Alfv\'{e}n speed,
which usually approximates the outflow jet speed. 
Another key parameter is the plasma beta,
the ratio of the plasma pressure to the magnetic pressure.
As will be discussed later, it corresponds to
the ratio of the Alfv\'{e}n velocity and the sound velocity.
The beta is typically below the unity
in the lower corona \citep{gary01,aschwanden} and
the lobe (upstream region) of
the magnetotail plasma sheet \citep{baum89}.

In this paper, we explore fine structures of a plasmoid in fast reconnection
when the upstream region consists of low-beta plasmas
by using two-dimensional MHD simulations. 
Employing a high-resolution shock-capturing (HRSC) code,
we successfully resolve a variety of shock structures. 
We confirm that
the new shock structures in Ref. \onlinecite{zeni10b} similarly appear
in the nonrelativistic regime. 
We further discuss the condition for these shocks, and
argue that they are common features of plasmoids in the low-beta plasmas. 
We also report some new features such as
the Kelvin-Helmholtz-like turbulence inside the plasmoid and
a potential signature of the corrugation instability of the slow shock surface.

This paper is organized as follows.
We describe our numerical setup in Sec. \ref{sec:setup}.
We present simulation results and then
investigate the local structures in depth
in Sec. \ref{sec:results}.
Section \ref{sec:beta} presents a brief parameter survey.
We discuss these results in Sec. \ref{sec:discussion},
particularly focusing on the plasma beta condition.
Section \ref{sec:summary} summarizes this paper.

\section{Numerical setup}
\label{sec:setup}

We employ the following resistive MHD equations.
\begin{eqnarray}
&&\frac{\partial \rho}{\partial t} + \div ( \rho \vec{v} )
= 0, \\
&&\frac{\partial \rho \vec{v}}{\partial t} + \div ( \rho\vec{v}\vec{v} + p_T\vec{I} - \vec{B}\vec{B} )
= 0, \\
&&\frac{\partial e}{\partial t}
+ \div \Big( (e+p_T )\vec{v} - (\vec{v}\cdot\vec{B}) \vec{B} + \eta \vec{j} \times \vec{B} \Big)
= 0, \\
&&\frac{\partial \vec{B}}{\partial t} + \div ( \vec{v}\vec{B} - \vec{B}\vec{v} ) 
+ \rot (\eta \vec{j})
= 0,
\end{eqnarray}
where $p_T=p+B^2/2$ is the total pressure
and $e=p/(\Gamma-1) + \rho v^2/2 + B^2/2$ is the total energy density. 
The polytropic index is set to $\Gamma=5/3$.
All other symbols have their standard meanings.

These equations are solved by a Godunov-type MHD code.
Numerical fluxes are calculated by
a Harten--Lax--van Leer (HLL) method.\cite{hll}
The spatial profiles are interpolated by
a monotonized central (MC) limiter.\cite{mc}
The electric currents at the cell surfaces are calculated
with the second order.\cite{toth08}
The second-order total variation diminishing (TVD) Runge--Kutta method
is used for the time-marching.\cite{shu88}
The hyperbolic divergence cleaning method
is employed for the solenoidal condition.\cite{dedner02}

The simulations are carried out in the $x$-$z$ two-dimensional plane.
The reconnection point is set to the origin $(x,z) = (0,0)$.
To reduce the computational cost, one quadrant ($x\ge 0, z\ge 0$)
is solved considering the symmetry of the reconnection system. 
We set a symmetric boundary at $x=200$ and
a conducting wall at $z=150$. 
The domain of $[0,200] \times [0,150]$ is
resolved by $6000 \times 4500$ grid cells. 
Harris sheet is employed as an initial model:
$\vec{B}(z) = \tanh(z) \vec{\hat{x}}$, $\vec{v}=0$,
$\rho(z) = 1 + \cosh^{-2}(z)/ \beta_{up}$, $p(z) = 0.5 ( \beta_{up} + \cosh^{-2}(z) )$,
where $\beta_{up}$ is the plasma beta in the upstream region.
Importantly, we consider that
the plasma beta is sufficiently lower than the unity, $\beta_{up} < 1$. 
We set $\beta_{up}=0.2$ in our main run. 
By definition
the Alfv\'{e}n velocity in the upstream region
is set to $c_{A,up} = |B|/\sqrt{\rho}=1$. 
Therefore the time is normalized by
the Alfv\'{e}n transit time of the Harris sheet thickness. 
Note that the evolution of reconnection is best
characterized by the Alfv\'{e}n transit time. 
The sound speed is uniform, $c_s = \sqrt{\Gamma \beta_{up} / 2} \sim 0.408$. 

It is known that the reconnection process is critically influenced
by the resistivity $\eta$.
In the MHD framework, a realistic fast reconnection requires
a localized diffusion region in a vicinity of the reconnection point\cite{petschek}.
This has been supported by simulations
with spatially-localized resistivity.\cite{ugai77,scho89,bis01}
Physically, such resistivity comes from local plasma instabilities,
Hall effects,\cite{birn01} or electron kinetic effects,\cite{hesse99}
none of which are fully understood.
It is also known that physics-minded anomalous resistivity models allow
fast reconnection.
The resistivity is enhanced by a phenomenological measure of microphysics,
e.g., the current intensity $(j/\rho)$.
Since such quantity is the strongest around the X-point,
the resistivity is localized and then
Petschek-type fast reconnection spontaneously switches on.\cite{ugai86,ugai92} 
Regarding the macroscopic plasmoid dynamics,
it was reported that both two models basically exhibit similar results.\cite{ugai95} 
In this work, to keep the system simple and
to see an evolution of a single reconnection,
we employ the localized model,
\begin{equation}
\label{eq:ohm}
\eta(x,z) = \eta_0 + (\eta_1-\eta_0)\cosh^{-2} ( \sqrt{x^2+z^2} ),
\end{equation}
where $\eta_1=1/60$ is the localized resistivity
and $\eta_0=1/1000$ is the background resistivity. 
A weak perturbation of $\delta A_y = -0.06 \exp[-(x^2+z^2)/4]$ is
imposed to magnetic fields to speed up the reconnection onset.

\section{Results}
\label{sec:results}

\begin{figure*}
\rotatebox{90}{\begin{minipage}{\textheight}
\begin{center}
  \ifjournal
  \includegraphics[width=\columnwidth]{f1.eps}
  \else
  \includegraphics[width=\columnwidth]{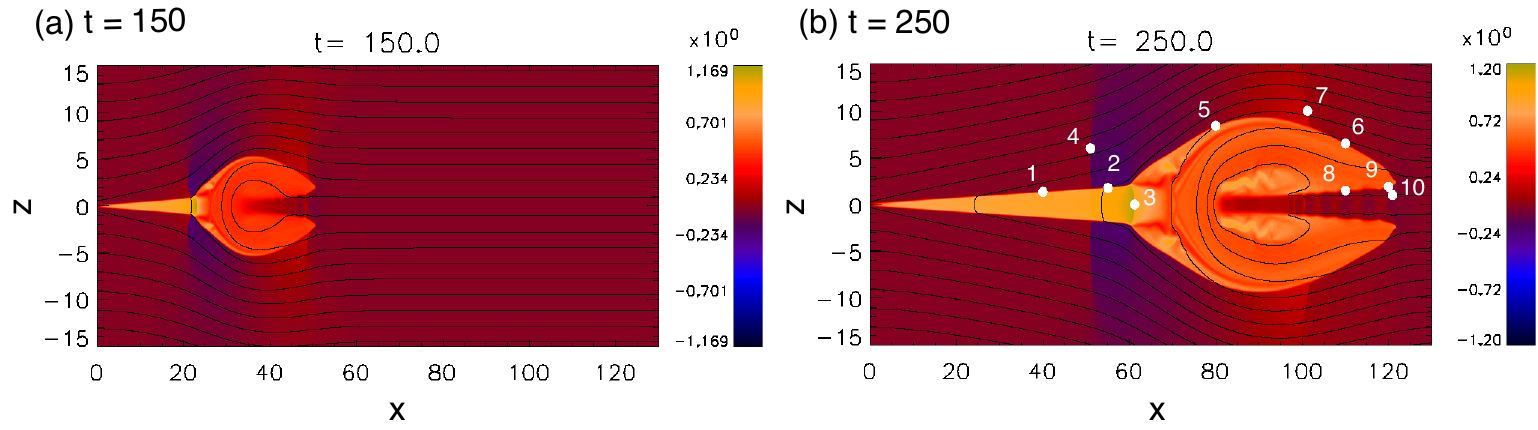}
  \fi
  \caption{
\label{fig:vx}
(Color online)
Time sequence of the velocity $v_x$, overlayed by the magnetic field lines.
The two panels stand for (a) at $t=150$ and (b) at $t=250$.
The white circles in the panel (b) indicates the locations
where we analysized shock parameters (see also Table \ref{table}).
(Enhanced online) 
}
\end{center}
\end{minipage}}
\end{figure*}

The panels in Fig. \ref{fig:vx} present the time development of $v_x$,
which most significantly characterizes the system evolution.
While we calculated the first quadrant ($z \ge 0$),
the bottom half ($z<0$) is inserted to understand the structures better. 
The evolution can also be seen in the movie available online.
The reconnection process starts around the origin. 
The localized resistivity causes magnetic field dissipation there,
giving rise plasma inflow from the upstream region. 
Reconnection transfers the upstream magnetic fields
toward the $+x$-direction along with an outflow jet. 
The flux transfer becomes faster and faster.
Around $t \gtrsim 100$,
the transfer rate $E_{y,R} = \eta_1 j_{y,R}$
reaches its typical value of $0.05 ~[c_{A,up}B_0]$,
where the subscript $R$ denotes the properties at the reconnection point. 
After that, the system exhibits a quasisteady evolution,
whose reconnection speed remains fast, $E_{y,R}\sim 0.05$. 
Using properties at an upstream position $(x,z)=(0,5)$ (denoted by $'$),
the rate can be re-normalized to $\mathcal{R} = E_{y,R}/(c'_{A}B') \sim 0.08$.
This is a fast reconnection rate of $\mathcal{R} \sim \mathcal{O}(0.1)$. 

Several characteristic structures already appear in Fig. \ref{fig:vx}a.
The reconnection jet becomes Alfv\'{e}nic ($v_{jet}\sim 1$) and
there is a small magnetic island (plasmoid) in front of the jet.
The plasmoid is heart-shaped\cite{saito95,hirai09},
because MHD velocities are faster in the low-dense upstream region
than ones in the plasma sheet ($z\sim 0$),
and because dense plasmas throttle the outward motion of the plasmoid
near the plasma sheet.
One can recognize vertical structures of velocity jumps
around $x\sim 21$ and $x\sim 48$ in the upstream plasmas. 
As the time goes on, the plasmoid grows bigger and bigger
as it travels to the $+x$-direction. 
It seems that
the macroscopic picture evolves self-similarly in later stages.\cite{nitta01}
This is because
the resistive spot and the Harris-sheet thickness become
relatively smaller than the entire system, as the system grows larger.
Meanwhile, fine structures become clear inside the plasmoid. 
Figure \ref{fig:vx}b shows the profile at $t=250$ of our main interest.
The vertical jump structures are found around $x\sim 60$ and $x\sim 100$.
The heart-shaped plasmoid now looks like a lobster-claw,
as the bifurcated parts become much larger than ones in the earlier stage.

Shown in Fig. \ref{fig:snap}a is
the plasma density $\rho$ and the flow pattern at $t=250$.
Many plasmas are confined in the plasmoid. 
Some of the upstream plasmas are swept into the plasmoid by the reconnection jet.
Other upstream plasmas directly enter the plasmoid across its outer boundary.
In addition, the plasmoid eats Harris-sheet plasmas while it moves to the $+x$-direction.
Figure \ref{fig:snap}b presents two thermodynamic properties.
The upper half shows the temperature including the Boltzmann constant $T=p/\rho$.
In general, plasmas are hot inside the reconnection jet and the plasmoid. 
The bottom half shows a useful measure $s=p/\rho^{\Gamma}$,
which behaves similarly as specific entropy.
Unless there is nonadiabatic heating,
it is conserved throughout the convection
$d/dt \big( {p}/{\rho^{\Gamma}} \big) = 0$.

In the Secs. \ref{sec:PK}-\ref{sec:front},
we take a closer look at local plasmoid structures at $t=250$,
from the left Petschek outflow to the front side of the plasmoid.
Note that boundary effects are completely ruled out.
Since the fastest magnetosonic speed is
$[ c^2_{A,up} + c^2_{s} ]^{1/2} \sim 1.08$,
even the initial impact has not yet come back
to the presented simulation domain.

\subsection{Petschek outflow}
\label{sec:PK}

As presented in Figs. \ref{fig:vx} and \ref{fig:snap}a, 
the fast outflow jet flows from the reconnection point. 
The jet is separated from the upstream regions
by slow shocks \citep{petschek}, and therefore
one can see hot and high-entropy plasmas inside the outflow channel
(Fig. \ref{fig:snap}b). 
Figure \ref{fig:1Dcut} presents one-dimensional properties
along the outflow line, $z=0$.
The jet speed quickly increases to $\sim 0.8$-$0.9$,
which is consistent with the Alfv\'{e}n speeds
in the upstream sides of the Petschek shocks. 
Note that the local Alfv\'{e}n speeds are slightly slower than
the initial speed of $1$, because the upstream conditions evolved. 
The jet speed increases again in the downstream ($x > 51$), but
we will discuss the downstream region later in Sec. \ref{sec:acc}.

To examine the shock properties in depth,
we carry out Rankine--Hugoniot analysis
in various locations in the simulation domain at $t=250$.
They are indicated by white marks in Fig. \ref{fig:vx}b
and the relevant results are presented in Table \ref{table}.
We determine the shock normal $\vec{\hat{n}}$ by
the minimum variance analysis \cite{son67}
and check the results by considering the magnetic coplanarity \cite{col66}.
We determine the shock velocity $v_{sh}$ by the mass conservation.
Then, we compare the normal flow speed in the shock frame
$(\vec{v}\cdot\vec{\hat{n}}-v_{sh})$
with the MHD wave speeds in the $\vec{\hat{n}}$-direction.
We calculate Mach numbers with respect to
the fast-mode, the Alfv\'{e}n (intermediate) mode,
and the slow-mode speeds:
$\mathcal{M}_f$, $\mathcal{M}_i$, and $\mathcal{M}_s$.
The subscript $1$ denotes the upstream properties and
$2$ denotes the downstream properties.
The first data (No. 1) in Table \ref{table} correspond to
the properties across the Petschek slow shock at $(x,z) \sim (40, \pm 1.35)$.
The slope angle of the Petschek shock is small, ${\sim}0.03$.
It is reasonable that the angle is a fraction of
the field line slope in the upstream side ${\sim}0.095$ or
the normalized reconnection rate $\mathcal{R}\sim 0.08$.
The shock is stationary, $v_{sh}=0$, as expected.
Across the shock,
only the slow-mode Mach number changes supersonic to subsonic (Table \ref{table}). 
This is a reasonable indication of the slow shock.

Inside the outflow,
we notice that there is a weak and narrow density cavity
along the neutral line $z=0$,
in the closer vicinity of the reconnection point.
The small box inside Fig. \ref{fig:snap}a zooms up the relevant region.
At $x=20$, the central density is $10\%$ lower than the surrounding density.
This is because plasmas are hotter there,
as pointed by a previous work \cite{shuei01}. 
The central plasmas are Joule-heated near the reconnection point,
while the surrounding plasmas are heated by the Petschek slow shocks. 
These two may be separated by contact discontinuities
in an idealized limit\cite{hoshino00,shuei01}.
Since our numerical scheme (HLL scheme) is weakly diffusive,
the boundaries become unclear further downstream. 
In the real world, the contact discontinuities will disappear
due to the field-aligned plasma mixing.

\subsection{Postplasmoid slow shock}
\label{sec:shock}

\begin{figure}
\begin{center}
  \ifjournal
  \includegraphics[width=\columnwidth]{f2.eps}
  \else
  \includegraphics[width=\columnwidth]{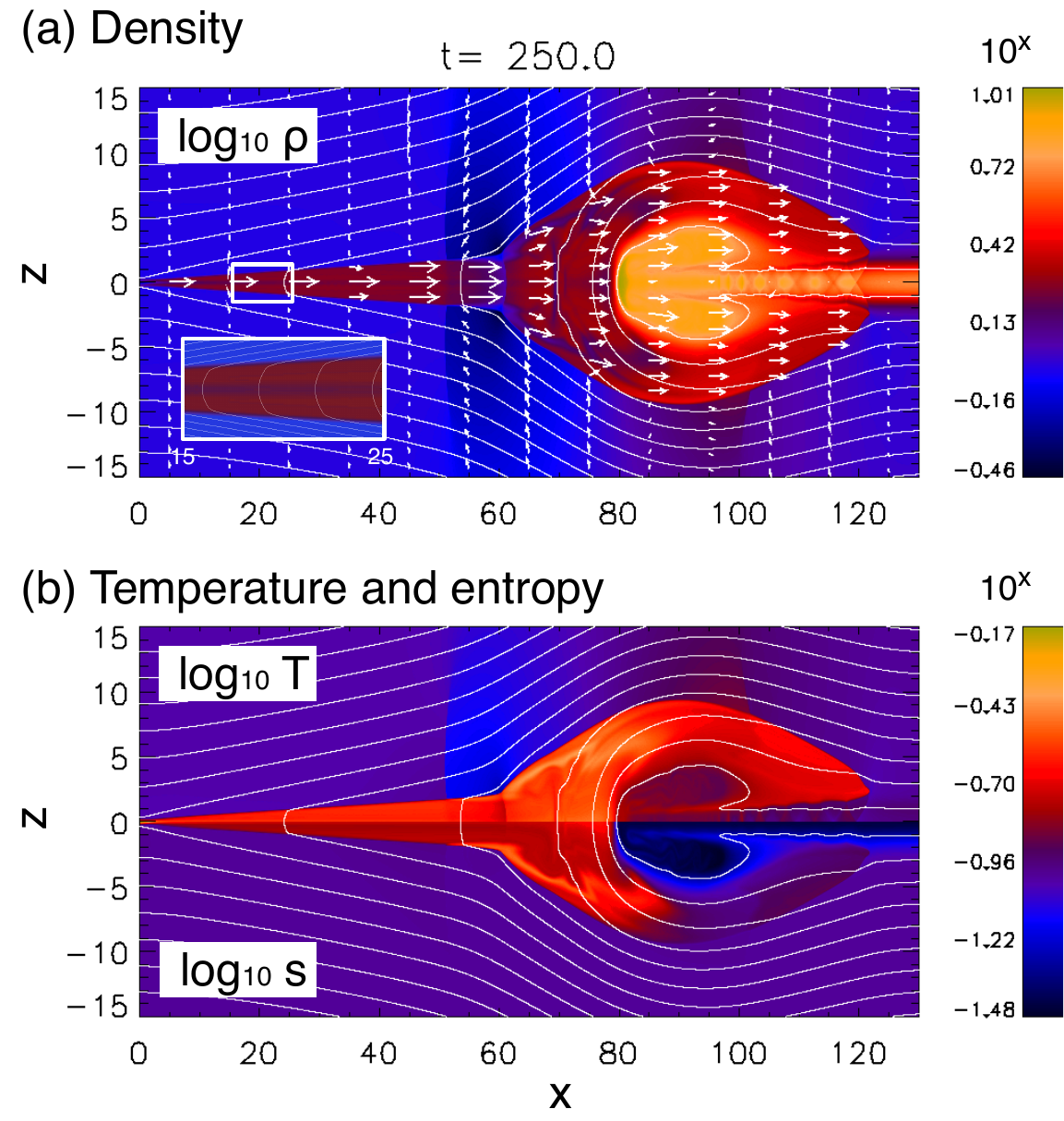}
  \fi
  \caption{
\label{fig:snap}
(Color online)
Snapshots at $t=250$, overlayed by magnetic field lines.
(a) Plasma density $\rho$.
Arrows indicate plasma velocity:
$|v|=1$ is projected to the length of $5$.
The inner box shows the smaller doamin of $x \in [15,25]$ and $z \in [-1.5,1.5]$.
(b) Plasma temperature $T$ in the upper half and
the entropy measure $s=p/\rho^{\Gamma}$ in the bottom half.
}
\end{center}
\end{figure}

Behind the plasmoid
one can see a sharp vertical structure at $x\sim 51$
in $v_x$ and $T$ (Figs. \ref{fig:vx} and \ref{fig:snap}b). 
We find that this is a slow shock.
We examine the shock properties near
a typical point of $(x,z) \sim (51, 6)$. 
The Rankine--Hugoniot properties are presented in No. 4 in Table \ref{table}. 
To better understand the structure,
one-dimensional properties along $z=6$ is
also presented in Fig. \ref{fig:RH}. 
Note that the shock normal is virtually parallel to the $x$-axis. 
The normal and tangential components are
similar to the $x$ and $z$-components, respectively. 
In this case, the right side is the shock upstream,
the left is the downstream, and
the shock moves to the $+x$-direction
at the speed of $v_{sh} \sim +0.31$. 
In Fig. \ref{fig:RH}a, the thin line shows
the normal flow speed in the shock frame, $v'_n = v_n - v_{sh}$. 
For comparison, the local-frame slow-mode speed
in the normal direction is presented as the dotted line. 
Comparing them and the slow-mode Mach numbers $\mathcal{M}_s$
in Table \ref{table},
one can see that
an upstream supersonic flow in the right side
becomes subsonic in the left downstream. 
The shock exhibits many other slow-shock features.
Along with the compression (Fig. \ref{fig:RH}b) and
the adiabatic heating ($T$ in Fig. \ref{fig:snap}b),
one can see that
the tangential magnetic field $B_t$ ($\approx B_z + 0.03$)
decreases across the shock (Fig. \ref{fig:RH}b). 
Meanwhile, this is a nearly parallel shock and
magnetic dissipation is quite limited. 
The measure $s$ is virtually unchanged (Fig. \ref{fig:snap}b) and
the heating is weak ($T_2/T_1$ in Table \ref{table}).

As discussed in Ref. \onlinecite{zeni10b},
this vertical shock is related to a reverse plasma flow around the plasmoid.
When the plasmoid moves to the right,
it compresses plasmas in the nearby upstream region.
Consequently, gas pressure drives plasma flows in the field-aligned directions.
In addition, since there is a density cavity behind the plasmoid (Fig. \ref{fig:snap}a),
the reverse flow in the postplasmoid region ($50<x<70$) is particularly strong,
as can be seen in Fig. \ref{fig:vx}b ($v_x<0$; blue regions). 
The reconnection inflows usually move to the $\pm z$-directions, but
they are accordingly modulated by the reverse flows (Fig. \ref{fig:snap}a).
Due to the field-aligned expansion,
the temperature becomes lower in the reverse flow region (Fig. \ref{fig:snap}b).
The slow shock is located at the front,
where the reverse flow supersonically hits the right-going shock front.

\subsection{Postplasmoid outflow}
\label{sec:acc}

\begin{center}
\begin{figure}
  \ifjournal
  \includegraphics[width=\columnwidth]{f3.eps}
  \else
  \includegraphics[width=\columnwidth]{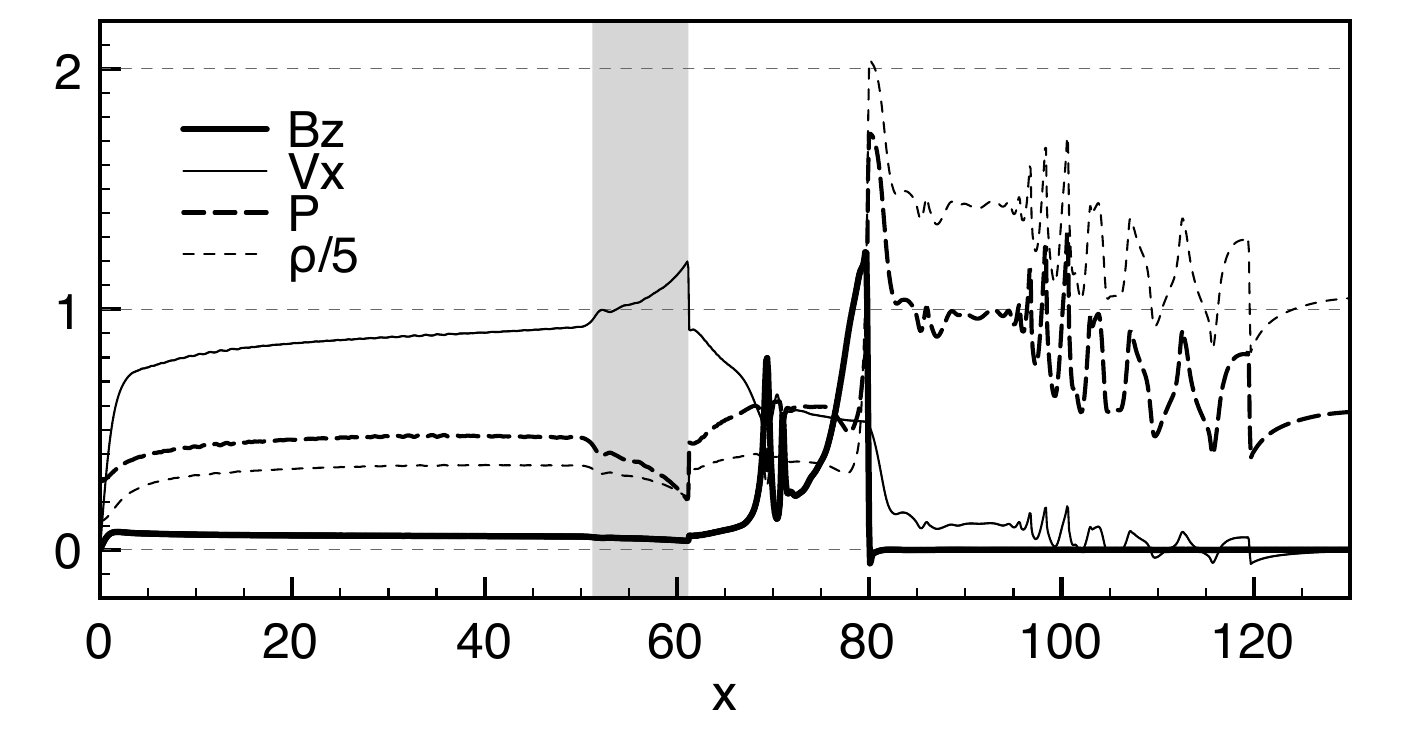}
  \fi
  \caption{
\label{fig:1Dcut}
One dimensional cut of physical properties along the neutral line, $z=0$.
The reconnected magnetic field $B_z$, outflow velocity $v_x$,
plasma pressure $p$, and the density $\rho$.  
The shadow indicates the region between
the vertical slow shock ($x=51.2$) and the fast shock ($x=61.2$).}
\end{figure}
\end{center}

\begin{center}
\begin{figure}
  \ifjournal
  \includegraphics[width=\columnwidth]{f4.eps}
  \else
  \includegraphics[width=\columnwidth]{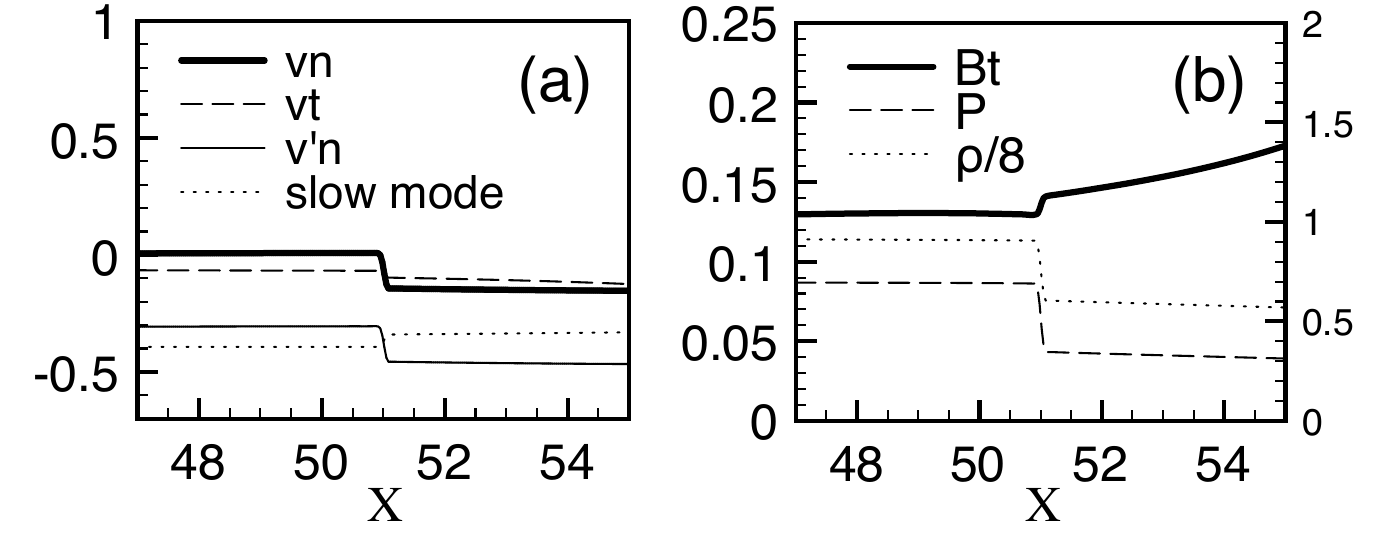}
  \fi
  \caption{
\label{fig:RH}
(a) One dimensional cut along $z=+6$.
Normal velocity $v_n$ ($\approx v_x$),
tangential velocity $v_t$, ($\approx v_z$),
shock-frame normal velocity $v'_n$, and
the shock-frame slow-mode velocity in the $-x$-direction.
(b) Tangential magnetic field $B_t$, pressure $p$, and density $\rho$.
See the second vertical axis to measure $\rho$.}
\end{figure}
\end{center}

We find that
the postplasmoid vertical shocks affect
the reconnection jet inside the outflow channel.
The vertical shocks meet Petschek shocks at $x \sim 51.2$.
In Fig. \ref{fig:1Dcut},
one can see that the reconnection jet suddenly becomes faster around there.
The jet further travels in the $+x$-direction, and then
it hits the plasmoid at a shock at $x\sim 61.2$. 
The results of the shock analysis are shown in No. 3 in Table \ref{table}.
This is a reverse-type fast shock \cite{forbes83,ugai87}. 
The shadow in Fig. \ref{fig:1Dcut} indicates the outflow region
between the vertical shock and the fast shock ($51.2 < x < 61.2$).
It seems that the pressure gradient accelerates the jet in this shadow region.
The jet speed reaches $v_x=1.2$ at the fast shock,
which is much faster than the initial upstream Alfv\'{e}n speed, $c_{A,up}=1$.
A similar effect is reported by \citet{shimizu00,shimizu03},
who claimed that deconfinement of the outflow channel leads to
an adiabatic-type fluid acceleration. 

We think that two mechanisms work in the shadow region.
First, the system adjusts the balance condition of the Petschek shock,
based on the upstream conditions.
In the left quiet region ($x < 51.2$),
plasmas from the downstream side of the vertical shock
come in to the Petschek shock.
The outflow speed is well approximated by
the upstream Alfv\'{e}n velocity there.
In the shadow region ($51.2 < x$),
plasmas from the upstream side of the vertical shock
come in to the Petschek shock.
The upstream Alfv\'{e}n speeds are faster there, $c_{A,up'}\sim 1.16$,
mainly due to the lower density (Fig. \ref{fig:RH}b).
Then, the Lorentz force at the Petschek shock accelerates plasmas
to the ``new'' outflow speed of ${\sim}c_{A,up'}$.
One can see a rarefaction-like transition between
the left Petschek flow and the right new Petschek flow
in Fig. \ref{fig:1Dcut}.
Second, an adiabatic-type acceleration further boosts the jet.\cite{shimizu00,shimizu03}
This is a simple hydrodynamic effect. 
Since the outflow speed exceeds the local sound speed of $c_s\sim 0.6$-$0.7$,
such a supersonic flow is adiabatically accelerated
when its cross section increases like a nozzle.
It particularly works near the fast shock,
where the plasmoid opens up the outflow channel,\cite{shimizu00,shimizu03}
and so the final outflow speed of $\approx 1.2$ exceeds
the maximum upstream Alfv\'{e}n speed of $c_{A,up'}\sim 1.16$.

According to the Rankine--Hugoniot analysis (No. 2 in Table \ref{table}),
the Petschek shock in the postplasmoid region
is stronger than the left-side (No. 1). 
The shocked plasmas are hotter and the entropy is higher
than plasmas which come from the left Petschek shock,
as seen in Fig. \ref{fig:snap}b.
A sharp boundary separates plasmas from the left and right Petschek shocks.
We think this boundary is a contact discontinuity,
which originates from the shock-shock interaction point.
Similarly, we expect that
the field-aligned mixing will blur the discontinuity in the real world.

\subsection{Outer shell and external slow shocks}
\label{sec:outer}

The plasmoid is surrounded by a relatively sharp boundary. 
The Rankine--Hugoniot analysis across this outer shell (Nos. 5 and 6 in Table \ref{table})
tells us that it is a slow shock \cite{ugai95}.
Judging from the heating ratio of $T_2/T_1$,
the shock is stronger in the back side of the plasmoid
than that in the front side,
as discussed in previous literature\cite{ugai95}.
In addition to the reconnection jet,
the slow shock on the outer shell supplies hot plasmas into the plasmoid.

Outside the plasmoid,
we find another pair of slow shocks
around $(x,z) \sim (100, \pm 10)$. 
The Rankine--Hugoniot result is presented in No. 7 in Table \ref{table}. 
This shock is rather weaker than the other shocks.
In fact, one can hardly find a jump in the measure $s$ in Fig. \ref{fig:snap}b,
because dissipation is quite limited. 
However, our code does capture
a clear jump structure in the velocity (Fig. \ref{fig:vx}b).
Interestingly, as can be seen in Fig. \ref{fig:vx}a and
the movie available online,
the shock is originally located near the front edge.
It gradually shifts backward to a stable location
as the system evolves.

\subsection{Internal structure}
\label{sec:internal}

The reconnection jet hits the plasmoid
at the fast shock (Sec. \ref{sec:acc}),
where plasmas enter the plasmoid.
In the downstream side,
there is a busy transition region at $x \sim 70$,
where the shocked flow further slows down and
accumulate the magnetic fields $B_z$ (Fig. \ref{fig:1Dcut}). 
This region corresponds to the loop-top front\cite{ugai87}.
Plasmas hit the loop-top front and then
they diverge from the central region to the upper and lower regions
inside the plasmoid. 
The reconnected fields are piled up between the loop-top front and
the sharp tangential discontinuity (TD) at $x \sim 80$. 
The field $B_z$ is remarkably strong in the left side of the TD.
The TD separates upstream-origin plasmas and
plasmas initially located inside the Harris current sheet.
This is visible in the $s$-profile in Fig. \ref{fig:snap}b.
The entropy is high in upstream-origin plasmas, because they are shock-heated or Joule-heated.
On the other hand, the right plasmas just come from the low-entropy Harris sheet.

Figure \ref{fig:plasmoid}a presents the density profile inside the plasmoid.
While the plasmoid is dynamically growing and expanding,
the TD travels in the $+x$-direction at a nearly constant speed of $v_{T} \approx +0.5$. 
To understand the plasma motion better,
the velocity profile in the moving frame of $v_{T}$ is shown by arrows.
Plasmas at $z\sim 0$ constantly travels in the $-x$-direction from the right. 
After they hit the TD, they turn to the $+z$ direction, and then
they are reflected to the $+x$-direction around $z\sim 2$.
The TD works as a ``magnetic wall'' to reflect plasmas from the right side.
In the original frame, this traveling wall hits
the immobile Harris-sheet plasmas ($v\sim 0$).
Therefore, the reflected flows are twice faster than
the wall motion in Fig. \ref{fig:vx}b, $\sim 2 v_{T} \sim 1$.

The fast reflected flow triggers velocity-shear-driven turbulence.
In the online movie, one can see that
a vortex rolls up where the flow turns its direction and
that the reflected flow is deformed to a kink-type structure. 
We think this is due to the Kelvin-Helmholtz instability
between the reflected jet and the surrounding medium. 
For example, typical parameters on the upper side of the reflected jet are
$B\sim B_0/2 \sim 0.5$, $\rho \sim \rho({t=0,z=0}) \sim 6$,
and $\Delta v \sim v_T \sim 0.5$, where $\Delta v$ is the velocity shear.
From the linear theory,\cite{miura82}
we find that
the shear is strong enough to bend the magnetic field lines,
$\Delta v > 2|B|/\sqrt{\rho}$,
but is slower than the sonic upper limit, $\Delta v < 2 c_s$.
Therefore, the shear layer would be unstable to the Kelvin-Helmholtz instability.
One can see that the reflected flow is violently flapping
in Fig. \ref{fig:vx}b and the online movie. 
The reflected flow and the resulting turbulence
mix plasmas inside the plasmoid.

In Fig. \ref{fig:plasmoid}a and the online movie,
one can also see that the incoming flow repeatedly flaps
in the right region of $98 \lesssim x$ and $z\sim 0$.
This is related to the shock-reflection structure
which will be discussed in Sec. \ref{sec:front}.

\subsection{Oblique-shock reflection}
\label{sec:front}

\begin{center}
\begin{figure}
  \ifjournal
  \includegraphics[width=\columnwidth]{f5.eps}
  \else
  \includegraphics[width=\columnwidth]{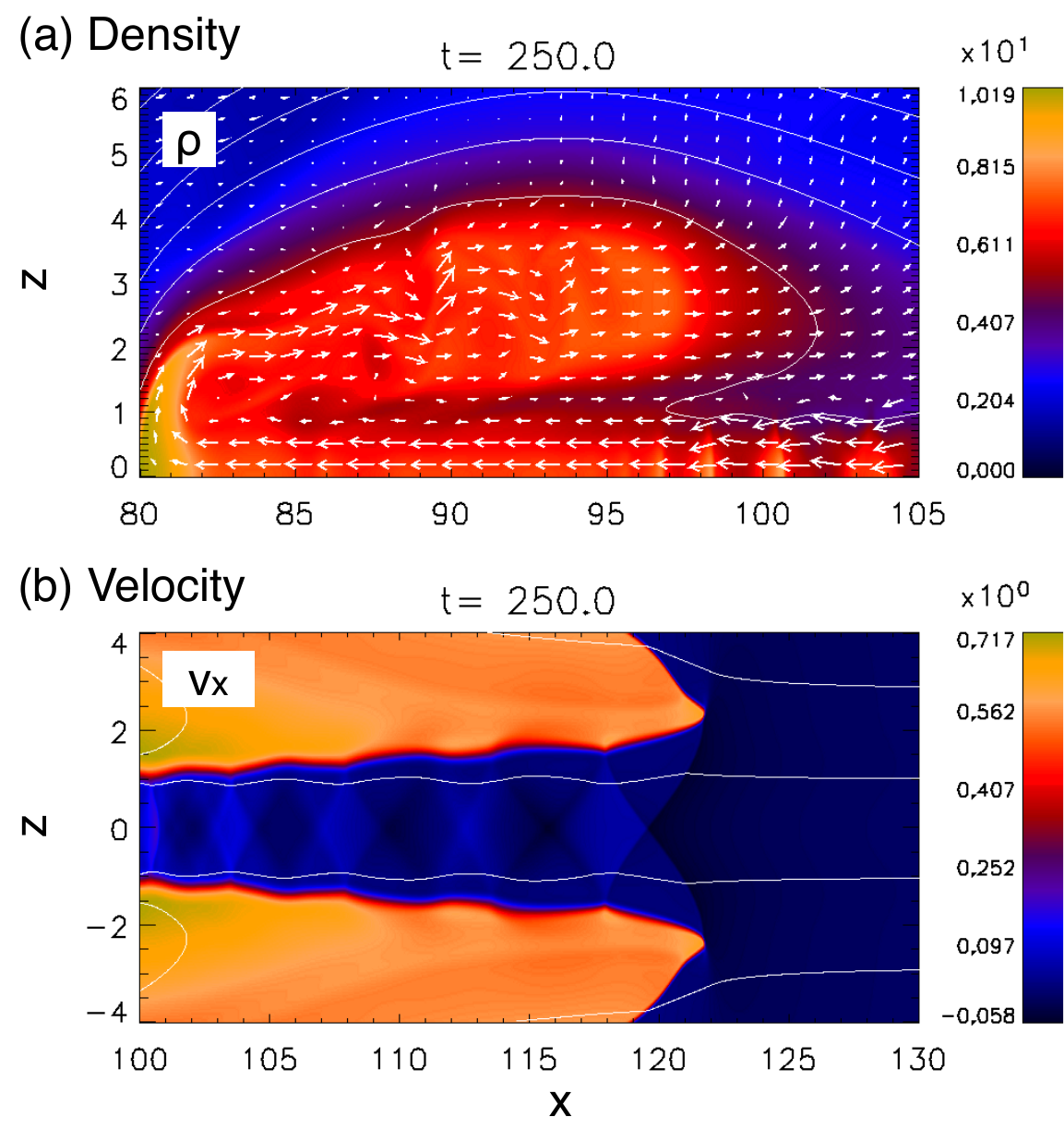}
  \fi
  \caption{
\label{fig:plasmoid}
(Color online)
(a) Plasma density $\rho$ inside the plasmoid at $t=250$.
The white lines indicate magnetic field lines.
The arrows indicate the plasma velocity
in the moving frame of $v_x=+0.5$.
The spatial length $2$ corresponds to $|v|=1$.
(b) Profile of $v_x$ at $t=250$ in the front side.
(Enhanced online) 
}
\end{figure}
\end{center}

Shown in Fig. \ref{fig:plasmoid}b is
the $v_x$-profile in the front side of the plasmoid.
The outer shells reach the leading edges of the plasmoid
at $(x,z) \sim (121.8, \pm 2.4)$.
Around the central region,
there is immobile plasmas ($v\sim 0$), because
high-density Harris-sheet plasmas are initially confined here.
Outside the center, plasmas fast travel in the +$x$ direction.
These twin flows are essentially driven by
the reconnection outflow or the reflection by the magnetic wall.
The central plasmas and the twin flows are
separated by transition regions.
The magnetic fields and plasmas flows are
rather parallel to the transition regions, but
they gradually enters from the central Harris sheet to the plasmoid flows.

As pointed out by \citet{shuei01},
the transition region looks like an intermediate shock,
across which the tangential magnetic fields change their polarities.
Unfortunately, our Rankine--Hugoniot analysis fails to tell
whether or not this is an intermediate shock
(No. 8 in Table \ref{table}).
This is probably due to the following three reasons. 
First, the assumption for the analysis may be invalid here.
The Rankine--Hugoniot analysis deals with a time-stationary and one-dimensional structure.
However, oblique shocks repeatedly hit the boundary regions as discussed later.
A time-dependent flow and a weak curvature makes our analysis difficult. 
Second, the slow and intermediate speeds are hard to separate
when the field lines are quasiperpendicular to the normal direction.
Third, our numerical scheme (HLL scheme) is not ideal for the structure here. 
The HLL code excellently captures the fast shocks but
handles other discontinuities in an {\itshape upwind} manner. 
When the characteristic speeds are slower than
the fast-mode speeds by order of magnitude,
the code smoothes the structure. 
Here, the fast speeds are typically $\sim 0.5$-$0.6$,
while intermediate and slow speeds are 
on an order of $\sim {\mathcal O}(0.01)~[c_{A,up}]$. 
These three reasons make it difficult to discuss the slow/intermediate Mach numbers.
In Table \ref{table}, we add asterisk signs to such unreliable Mach numbers.
The third issue will be improved by a better scheme
which fully considers intermediate discontinuities \cite{miyoshi05}.
In the right side near the flow edges (No. 9 in Table \ref{table}),
the boundaries look like slow shocks\cite{saito95}.
Unfortunately, we fail to classify this shock for the same reasons.
We expect that the intermediate shock changes to the slow shock
through the switch-off shock \cite{hau89,hirai09}.

In the central channel,
one can see that the oblique shocks propagate from the twin flow edges
(Fig. \ref{fig:plasmoid}b).
They are fast shocks (No. 10 in Table \ref{table}).
This is because the edges travel faster than
the fastest right-going wave inside the central channel. 
These oblique shocks are repeatedly reflected by
the intermediate shock regions \cite{shuei01}.
Such shock-reflection leads to
a chain of the diamond-shaped patterns,
as denoted by the ``diamond-chain'' structure in Ref. \onlinecite{zeni10b}. 
The second and third right ones look irregular in this case, but 
one can see that all other rectangles are excellently in order in Fig. \ref{fig:plasmoid}b. 
The repeated shock-reflection explains
a violent fluctuation in sonic properties
in the front region ($95 < x < 120$) in Fig. \ref{fig:1Dcut}.

Roughly speaking, the plasmoid edges travel
at a sizable fraction of the reconnection jet speed $v_{jet}$,
which is well approximated by the initial upstream Alfv\'{e}n speed, $c_{A,up}$.
On the other hand, since magnetic fields are very weak inside the current sheet,
the right-going fast-mode speed is approximated by
the sound speed in the current sheet.
Therefore, a requirement for the oblique shock and the diamond-chain is
\begin{eqnarray}
\label{eq:diamond}
{c_{s,cs}} < {c_{A,up}},
\end{eqnarray}
where the subscript $cs$ denotes the current sheet properties. 
Needless to say, our initial configuration satisfies this:
$c_{s,cs} = c_{s} = 0.408$ and $c_{A,up} = 1$.

In the fronst side,
there are small bow-shock-like structures around $x=121.8$, $z=2.55$--$2.75$,
extending outward from the plasmoid edges.
We do not recognize steep shocks there. 
We think they are the wave fronts of the fast mode \cite{saito95}.

\section{Beta Dependence}
\label{sec:beta}

\begin{center}
\begin{figure}
  \ifjournal
  \includegraphics[width=\columnwidth]{f6.eps}
  \else
  \includegraphics[width=\columnwidth]{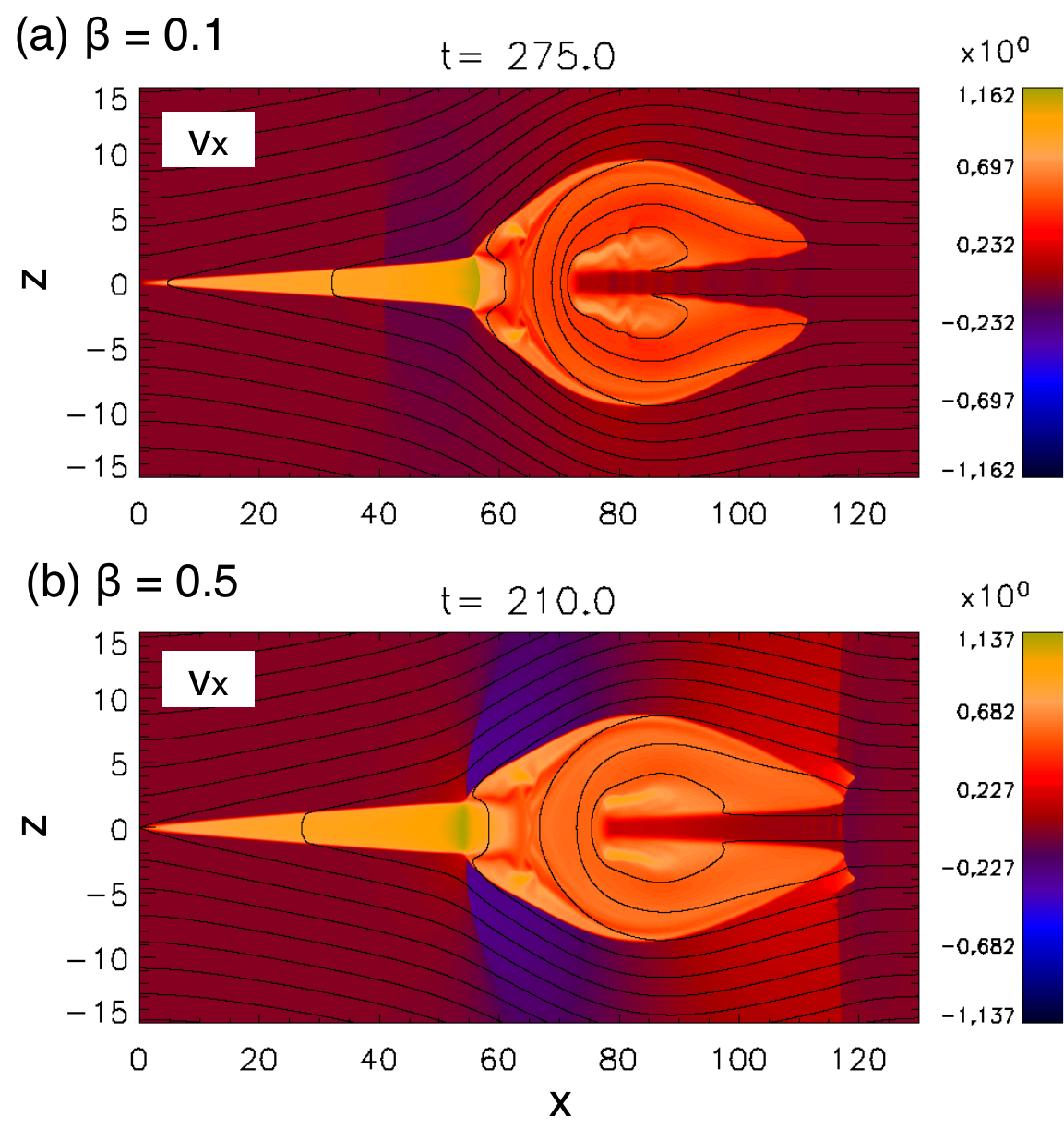}
  \fi
  \caption{
\label{fig:beta}
(Color online)
The $v_x$-profiles in different simulation runs:
(a) $\beta_{up}=0.1$, $t=275$ and (b) $\beta_{up}=0.5$, $t=210$.
}
\end{figure}
\end{center}

We carry out another two runs using
different $\beta_{up}$ parameters, $\beta_{up}=0.1$ and $0.5$.
All other parameters are the same as the main run.
Figure \ref{fig:beta}a represents a snapshot
with a lower upstream beta, $\beta_{up}=0.1$.
The system builds up slower than the main run,
because the system has relatively denser plasmas in the downstream Harris sheet. 
One can recognize many characteristic signatures except for
the forward vertical slow shocks outside the plasmoid.
Actually, there {\it are} slow shocks near $(x,z) \sim (82,\pm 9.8)$,
but they are too small to recognize in the figure. 
The postplasmoid vertical shocks are located at $x\sim 40$.
The velocity jump at the shock is smaller than the main run,
because the sound speed is initially slower,
$c_s = \sqrt{\Gamma \beta_{up} / 2} \sim 0.29$.

Figure \ref{fig:beta}b shows the other run with $\beta_{up}=0.5$.
One can see that the postplasmoid slow shocks are located
in the back side of the plasmoid's outer shell, $x \sim 55$.
The other pair of slow shocks are located near
the front edges of the plasmoid, $x \sim 115$.
At a glance the front shocks look like switch-on fast shocks,
but they are switch-off type slow shocks. 
Magnetic fields are gradually bent
in the right side of the forward vertical slow shocks.
Interestingly, a finger-like structure can be seen
on the slow shock surface,
around $2 < |z| < 6$ outside the plasmoid-front.
It slowly grows there and
it may eventually destroy and blur the forward slow shocks. 
We think that this is a signature of the corrugation instability of
the slow shock surface \cite{stone95}.
However, detail analysis of the instability is beyond the scope of this paper.

\section{Discussion}
\label{sec:discussion}

\begin{center}
\begin{figure}
  \ifjournal
  \includegraphics[width=\columnwidth]{f7.eps}
  \else
  \includegraphics[width=\columnwidth]{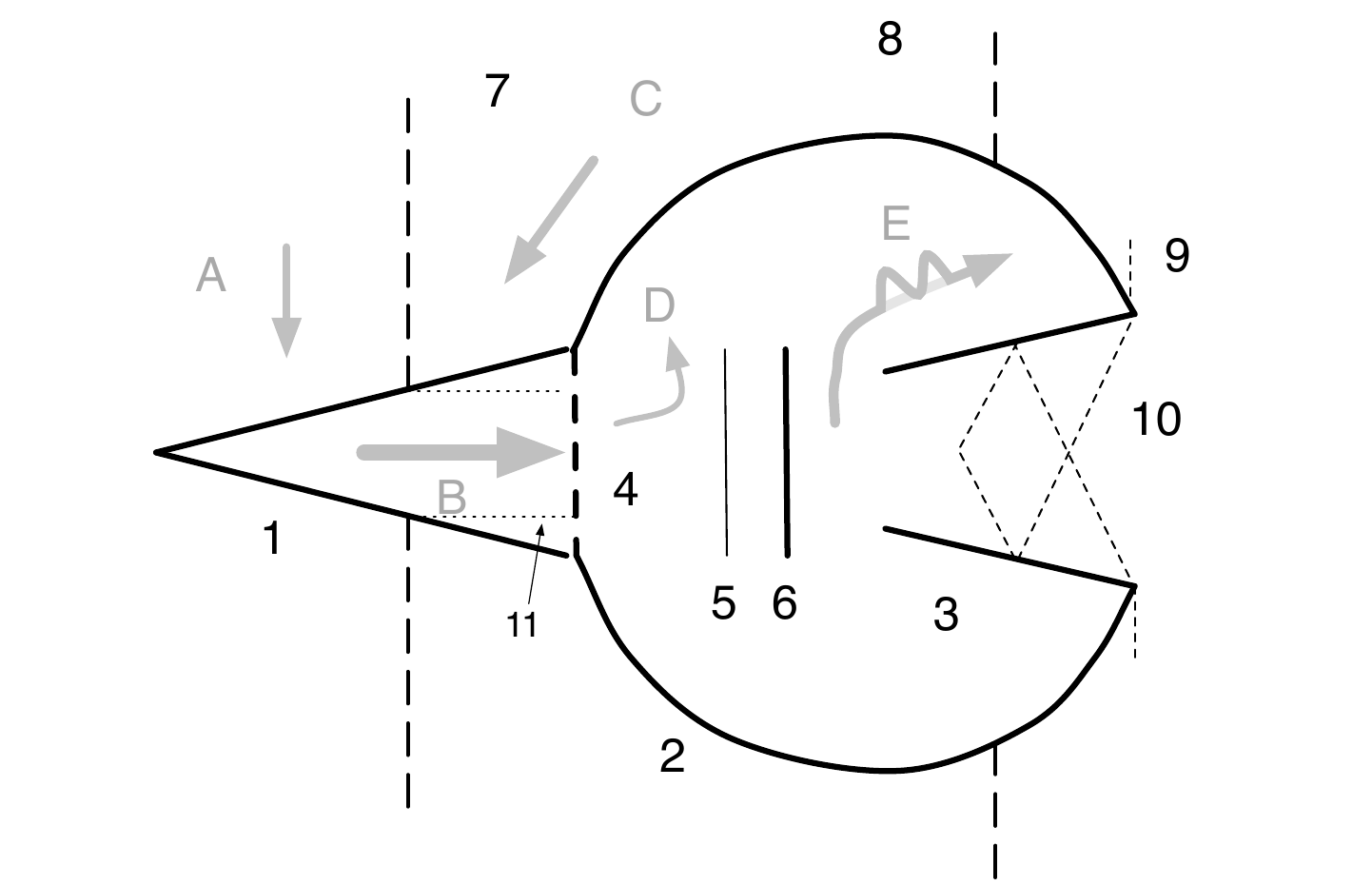}
  \fi
  \caption{\label{fig:illust}
Schematic drawing of plasmoid structures:
1. Petschek slow shock \cite{petschek},
2. outer shell (slow shock)\cite{ugai95},
3. intermediate shock\cite{shuei01} and slow shock\cite{saito95},
4. fast shock \cite{forbes83,ugai87},
5. loop-top front \cite{ugai87},
6. tangential discontinuity,
7. postplasmoid vertical slow shock \cite{zeni10b},
8. forward vertical slow shock (Sec. \ref{sec:outer}),
9. fast-mode wave front \cite{saito95},
10. shock-reflection (diamond-chain) \cite{zeni10b},
11. contact discontinuity (Sec. \ref{sec:acc}),
A. reconnection inflow,
B. outflow jet,
C. postplasmoid backward flow,
D. internal flow, and
E. flapping jet.
}
\end{figure}
\end{center}

In this paper we visited many structures in and around the plasmoid.
They are schematically illustrated in Fig. \ref{fig:illust}.
This extends an informative result in an earlier work
(i.e., Fig. 10 in Ref.~\onlinecite{shuei01}).
Among these structures,
the postplasmoid vertical shocks (Sec. \ref{sec:shock})
and the shock-reflected diamond-chain structure
in the front side (Sec. \ref{sec:front})
were recently found in the relativistic regime.\cite{zeni10b}
We confirmed them in the nonrelativistic regime as well ---
they are ubiquitous, regardless of the relativistic effects. 
A reverse flow or a backward flow around the plasmoid is often found
in recent large-scale simulations\cite{shuei01,zeni09a},
however, we clearly demonstrated that a slow shock stands at the front of the reverse flow.
We also introduced another vertical slow shock outside the plasmoid. 
Those two vertical shocks are similarly found
in other runs in our parameter survey. 

We think that
the vertical slow shocks are logical consequences of
the low-beta condition of $\beta_{up} \lesssim 1$.
From the following relation,
\begin{eqnarray}
\label{eq:beta}
( {c_{s,up}}/{c_{A,up}} )^2
= ({\Gamma}/{2}) \beta_{up}
,
\end{eqnarray}
we see 
$c_{s,up} < c_{A,up}$ when $\beta_{up} \lesssim 1$.
As already discussed, the plasmoid-front propagates
at a sizable fraction of the reconnection jet speed $v_{jet}\sim c_{A,up}$. 
When $\beta_{up} \lesssim 1$, the plasmoid supersonically travels in the system.
At some location, the expanding plasmoid structure interacts
with the surrounding plasmas at a supersonic speed,
and then a shock stands there.
Another interpretation is as follows.
In a self-similarly expanding coordinate of the plasmoid system,\cite{nitta01}
the surrounding plasmas supersonically flow in the $-x$-direction near the plasmoid-front,
while the relative flow is very small near the reconnection point.
A shock stands at the point
where the supersonic flow slows down to subsonic speed.
In summary, when $\beta_{up} \lesssim 1$,
the plasmoid system will inevitably involve vertical slow shocks.

In addition, the compression effects by the plasmoid introduces
another pair of vertical shocks outside the plasmoid.
As discussed in Sec. \ref{sec:shock},
the compression invokes the field-aligned flows in the ${\pm}x$ directions.
In addition, the adiabatic heating or cooling weakly modify the local sound speed,
$c_s = \sqrt{\Gamma s^{\frac{1}{\Gamma}}} \cdot p^{(\Gamma-1)/{2\Gamma}}
\propto p^{{1}/{5}}$.
In the main run,
the relative velocity between the plasmoid system and the upstream plasmas
are initially supersonic in the right side.
It temporally becomes subsonic at the forward vertical shock $x \sim 100$
(Sec. \ref{sec:outer}) due to the compression effects.
However, due to the field-aligned acceleration in the $-x$-direction,
the flow becomes supersonic again, and then
it becomes subsonic at the postplasmoid vertical shock
at $x\sim 51$ (Sec. \ref{sec:shock}).

The positions of the vertical shocks depend on the $\beta_{up}$ parameter,
because it controls the ratio of the sound speed to the Alfv\'{e}nic plasmoid motion.
When $\beta_{up}$ is lower than the main run ($\beta_{up}=0.1$), 
the expanding speed of the entire structure easily exceeds the sound speed.
Therefore, the postplasmoid vertical shock is closer to the reconnection point,
as demonstrated in Sec. \ref{sec:beta} (Fig.~\ref{fig:beta}a).
The forward vertical slow shock can be small,
when the plasmoid compression effects are not strong enough to
change the supersonic condition. 
Probably the forward vertical shock disappears
in the extreme limit of $\beta_{up} \ll 1$.
On the other hand,
when $\beta_{up}$ is higher ($\beta_{up} = 0.5$), 
the vertical slow shocks are nearer to the front-side of the plasmoid:
the postplasmoid and forward ones are located
just behind and in front of the plasmoid, respectively.
As $\beta_{up}$ increases, the postplasmoid slow shock moves frontward
even in the backward side of the plasmoid outer shell, and then
it will eventually disappear.
The forward slow shock will disappear
when the Alv\'{e}n speed is no longer supersonic,
$\beta_{up} \sim 2/\Gamma = 1.2$.
Depending on $\beta_{up} (\lesssim 1)$, 
we expect one or two pairs of vertical slow shocks around the plasmoid.

Our discoveries can readily be applied to
various numerical results in low-beta plasmas.
\cite{shuei01,tanuma07,zeni10b,yu11}
\citet{shuei01} carried out large-scale MHD simulation of
a plasmoid in a similar configuration. 
While they did not mention it in the text,
one can see a shock-like structure outside the plasmoid
in RUN 2 with $\beta_{up}=0.25$
(the top panel in Fig. 9 in Ref. \onlinecite{shuei01}). 
We think that this is a forward vertical slow shock. 
\citet{zeni10b} recognized a small shock outside the plasmoid
(indicated by an arrow in Fig. 2c in Ref. \onlinecite{zeni10b})
in the relativistic regime with $\beta_{up}=0.1$.
Inspired by this work,
we carry out further analysis and confirm that
it was a forward vertical slow shock. 
In multiple plasmoid systems,
\citet{tanuma07} found ``bow shocks''
during the nonsteady reconnection
under an anomalous-type resistivity with $\beta_{up}=0.2$
(Fig. 1d in Ref. \onlinecite{tanuma07}). 
Judging from displacements,
their plasmoid velocities are sub-Alfv\'{e}nic,
on an order of ${\sim}c_{s,up} < c_{A,up}$. 
In such a developing stage, the forward vertical shock is located
near the front, and so it would be a forward slow shock.
Finally, a recent work by \citet{yu11} reported
a steepening structure of slow-mode wave
behind the primary plasmoid (Fig. 15a in Ref.~\onlinecite{yu11}).
It would be related to a postplasmoid slow shock,
as we found it in our relevant run with $\beta_{up} =0.5$
(Sec. \ref{sec:beta}; Fig. \ref{fig:beta}b).

The shock-reflection or the diamond-chain structure is
also a characteristic in the low-beta condition.
Dropping a factor of order unity,
the requirement (Eq. \ref{eq:diamond}) reads
\begin{eqnarray}
\Big( \frac{c_{s,cs}}{c_{A,up}} \Big) \sim
\Big( \frac{c_{s,cs}}{c_{s,up}} \Big)
\sqrt{ \beta_{up} } < 1.
\end{eqnarray}
Although we still have a free parameter $( {c_{s,cs}}/{c_{s,up}} )$,
the condition can be easily satisfied in the low-$\beta_{up}$ regimes.

Let us discuss some limitations of our simulation model.
First, we assumed that
the evolution in the bottom quadrant ($z<0$) is symmetric to
that in the upper quadrant ($z>0$).
We do not think the quadrant assumption affects the main shock structures,
because Ref. \onlinecite{zeni10b} was carried out without the assumption.
On the other hand, we found
the Kelvin-Helmholtz-like turbulence in the internal region (Sec. \ref{sec:internal}).
Since this class of instabilities may prefer odd-parity perturbation across
the neural line ($z=0$), the central Harris sheet may flap in very late stages.
This is left for future investigation. 
Second, in the three dimensions, it is known that
quasiperpendicular slow shocks are unstable to
the corrugation instability.\cite{stone95}
We do see a potential signature of the instability
even in the two dimensions (Sec. \ref{sec:beta}).
In the real world, the vertical slow shocks may be violently folded and then
we will see turbulent transition layers instead. 
Regarding the shock-reflected diamond-chain structure,
the oblique fast shocks will be stable\cite{gardner64}, but
it is not clear how stable the intermediate shocks are.
It is interesting to see how those structures are modulated in the three dimensions in future.

\section{Summary}
\label{sec:summary}

We carried out HRSC MHD simulations to study the fine structures of a plasmoid.
We introduced new shock structures: 
(1) postplasmoid vertical slow shocks, (2) forward vertical slow shocks, and
(3) shock-reflection structures,
some of which were recently reported in Ref. \onlinecite{zeni10b}.
We further found new features such as
(4) the flapping motion of the reflected jet and
(5) the finger-like instability of the slow shock surface.
We interpreted that the vertical slow shocks and the shock-reflection
are consequences of the Alfv\'{e}nic plasmoid motion,
which is supersonic in low-beta plasmas. 
We argue that these shocks are rigorous features of a reconnection system
in low-beta plasmas.

\begin{acknowledgments}
The authors are grateful to A.~F. Vinas for extensive discussion on shock analysis,
J.~C. Dorelli and A. Glocer for many advices on the code development,
S.~A. Abe, C.~R. DeVore, M. Hesse, M. Hoshino, S. Inutsuka, J. T. Karpen,
Y. Matsumoto, M. N. Nishino, and T. Shimizu for general discussions. 

S.Z. gratefully acknowledges support from JSPS Fellowship for Research Abroad. 
T.M. was partially supported by Grant-in-Aid for Young Scientists (B) (Grant No. 21740399).
\end{acknowledgments}

\clearpage

\begin{table*}
\begin{center}
\rotatebox{90}{\begin{minipage}{\textheight}
\begin{tabular}{lccccccccccccl}
\hline
\hline
\# & 
Location &
$(n_x,n_z)$ &
$v_{sh}$ &
$|\theta_{BN}|$ &
$\beta_1$ &
$\mathcal{M}_{f1}$ &
$\mathcal{M}_{i1}$ &
$\mathcal{M}_{s1}$ &
$\mathcal{M}_{f2}$ &
$\mathcal{M}_{i2}$ &
$\mathcal{M}_{s2}$ &
$T_2/T_1$ &
\\
\hline
1 &
(40.0, 1.35) &
(-0.03, 1.00) &
0.0 &
86.3 &
0.22 &
0.06 &
0.98 &
2.49 &
0.04 &
0.69 &
0.69 &
2.72 &
(S) Petschek shock
\\
2 &
(55.0, 1.75) &
(-0.04, 1.00) &
-0.013 &
86.3 &
0.098 &
0.06 &
0.88 &
3.22 &
0.04 &
0.58 &
0.58 &
4.58 &
(S) Petschek shock
\\
3 &
(61.2, 0) &
(-1.00, 0.00) &
-0.40 &
90 &
303 &
1.41 &
&
&
0.77 &
&
&
1.38 &
(F) Reverse shock
\\
4 &
(51.0, 6.0) &
(1.00,-0.04) &
0.31 &
9.4 &
0.12 &
0.41 &
0.42 &
1.34 &
0.33 &
0.34 &
0.78 &
1.33 &
(S) Postplasmoid vertical shock
\\
5 &
(80.0, 8.4) &
(-0.18, 0.98) &
-0.06 &
86.5 &
0.16 &
0.05 &
0.85 &
2.47 &
0.03 &
0.56 &
0.65 &
2.54 &
(S) Outer shell
\\
6 &
(110.0, 6.5) &
(0.24, 0.97) &
0.19 &
84.9 &
0.21 &
0.06 &
0.76 &
1.99 &
0.05 &
0.53 &
0.64 &
2.06 &
(S) Outer shell
\\
7 &
(101.2, 10.0) &
(0.94, 0.33) &
0.54 &
25.2 &
0.23 &
0.43 &
0.49 &
1.15 &
0.39 &
0.44 &
0.87 &
1.15 &
(S) Forward vertical shock
\\
8 &
(110.0, 1.5) &
(-0.06,-1.00) &
0.10 &
87.8 &
1.1 &
0.12 &
4.5* &
6.5* &
0.12 &
3.9* &
4.0* &
1.55 &
(U) Intermediate shock?
\\
9 &
(120.0, 1.9) &
(0.13,-0.99) &
0.13 &
87.1 &
0.49 &
0.09 &
2.0* &
3.8* &
0.08 &
1.7* &
1.9* &
1.86 &
(U) Slow shock?
\\
10 &
(120.9, 1.0) &
(0.64,-0.77) &
0.50 &
46.8 &
2.63 &
1.22 &
3.00 &
3.40 &
0.88 &
2.66 &
3.06 &
1.18 &
(F) Oblique shock
\\
\hline
\hline
\end{tabular}
\caption{\label{table}
Rankine--Hugoniot analysis. 
The subscripts $1$ and $2$ denote the upstream and downstream quantities.
The locations $(x,z)$ in the simulation domain (see also Figure \ref{fig:vx}b),
the shock normal vector $\vec{\hat{n}}$, the shock velocity $v_{sh}$,
the angle between $\vec{\hat{n}}$ and the upstream magnetic field $\vec{B}_1$,
the upstream plasma beta,
flow Mach numbers to fast, intermediate (Alfv\'{e}n), and slow-mode speeds, and
the temperature ratio.
The asterisk sign (*) indicates unreliable results (see Sec. \ref{sec:front}).
The letter (S) indicates a slow shock, (F) is a fast shock, and
(U) is unclassified.
}
\end{minipage}}
\end{center}
\end{table*}

\end{document}